\documentclass[prd,showpacs,showkeys,floatfix,nofootinbib,fleqn,tightenlines,
preprint]{revtex4}

\usepackage{amsmath}\usepackage{amsfonts}
\usepackage{latexsym}\usepackage{graphicx}
\usepackage{mathrsfs}\usepackage{dcolumn}

\newcommand{\version}{v3}               
\newcommand{\beq}{\begin{equation}}
\newcommand{\eeq}{\end{equation}}
\newcommand{\beqa}{\begin{eqnarray}}
\newcommand{\eeqa}{\end{eqnarray}}
\newcommand{\bsubeqs}{\begin{subequations}}
\newcommand{\esubeqs}{\end{subequations}}

\newcommand{\FPStype}{FPS}

\begin{document}
%
%
\noindent arXiv:1109.5671  \hfill KA--TP--26--2011\;(\version)\vspace*{2\baselineskip}%
\title{Superluminal muon-neutrino velocity from a
Fermi-point-splitting model of Lorentz violation}

\author{F.R. Klinkhamer}
\email{frans.klinkhamer@kit.edu}
\affiliation{
\mbox{Institute for Theoretical Physics, University of Karlsruhe,}
\mbox{Karlsruhe Institute of Technology, 76128 Karlsruhe, Germany}\\}

\begin{abstract}
\vspace*{.5\baselineskip}\noindent  
A particular Fermi-point-splitting (FPS) model is proposed,
which allows, for finite values of the three-momentum,
the group and phase velocities of the muon-flavor neutrino in vacuum
to be superluminal (i.e., muon-neutrino velocities greater than $c$,
the velocity of light in vacuum).
Still, the proposed FPS model has a front velocity
equal to $c$ for all particles, which is known to be
crucial for causality.
If the muon-neutrino velocity is indeed superluminal as
claimed by OPERA, the model predicts that
at least one of the other two flavors of neutrinos
has a subluminal velocity.
\vspace*{2\baselineskip}
\end{abstract}

\maketitle

The OPERA collaboration~\cite{OPERA2011}
has presented the following result
(combining the quoted statistical and systematics errors
in quadrature):
\beq\label{eq:OPERA-result}
\Big(
v_{\nu_\mu}\,
\Big|_{\langle c\,|\vec{p}| \rangle\, = \,17\;\text{GeV}}^\text{(exp)}
\,- c\,\Big)\Big/c = (+2.5\pm 0.4) \times 10^{-5}\,,
\eeq
with $c$ the velocity of light \textit{in vacuo}
and $v$
a particular velocity (time-of-flight) obtained
by a particular experimental procedure. Of course,
unknown systematic effects cannot be ruled out at this moment
(and are by many believed to be responsible for the claimed result),
but we will for now take the positive value
from \eqref{eq:OPERA-result} as a possible
experimental clue to radically new physics in the neutrino
sector.\footnote{An earlier result by MINOS~\cite{MINOS2007}
was perhaps suggestive but certainly inconclusive.
Still, the preferred values from MINOS and OPERA lie in the same
ballpark, $\big(v_{\nu_\mu}- c\big)\big/c$ ranging from
approximately $+10^{-5}$ to approximately $+10^{-4}$.}

Incidentally, the claimed result \eqref{eq:OPERA-result}
presents us with \emph{two} surprises.
The first surprise is that it would apply to neutrino propagation
essentially \textit{in vacuo}, without apparent dependence on the
neutrino energy~\cite{OPERA2011}.
The second surprise is that the magnitude of \eqref{eq:OPERA-result}
is very much above what may have been expected.
The expected value is around $10^{-19}$,
as follows, for example, from an indirect experimental bound
for a model with a universal velocity of the
standard-model gauge bosons and a universal maximum
attainable velocity of the fermions (see App.~C in
Ref.~\cite{KlinkhamerSchreck2008} for details).
In fact, Cohen and Glashow~\cite{CohenGlashow2011} have shown
recently that OPERA's $10^{-5}$ value is ruled out by the expected
but unobserved energy losses from electron-positron-pair emission,
$\nu_\mu\to \nu_\mu+e^{-}+e^{+}$.

The main goal of this Letter, now, is to show that a result
such as \eqref{eq:OPERA-result} is,
\textit{a priori}, not impossible in a Lorentz-violating (LV)
theory with universal maximal velocity $c$.
Quantitatively, we keep the experimental value
open by defining
\beq\label{eq:def-delta-nu-mu}
\delta_{\nu_\mu,\,L}^\text{\,(exp)}\big(|\vec{p}|\big)\equiv
\Big(v^\text{(group)}_{\nu_\mu,\,L} \big(|\vec{p}|\big)\Big/c
-1\,\Big)^\text{(exp)}\geq 0\,,
\eeq
with a relevant momentum value of $|\vec{p}|\sim 17\;\text{GeV}/c$
and assuming the parameter to be nonnegative
(i.e., superluminal or luminal muon-neutrino group velocity).
As discussed in the previous paragraph, it may very well be that
$|\delta_{\nu_\mu,\,L}^\text{\,(exp)}|\lesssim 10^{-19}$.

The particular class of theories considered is motivated
by condensed-matter physics and has as its crucial ingredient
the phenomenon of Fermi-point splitting (FPS) from a
quantum phase transition~\cite{KlinkhamerVolovikIJMPA2005};
see Ref.~\cite{Volovik2007} for a comprehensive review of
topology in momentum space.
It was immediately realized that these LV theories may
lead to new effects in neutrino
physics~\cite{KlinkhamerVolovikIJMPA2005,
KlinkhamerJETPL2004-IJMPA2006,KlinkhamerPRD2005-PRD2006};
see Ref.~\cite{Klinkhamer2006-review} for a brief summary.
These new effects (e.g., nonstandard \mbox{T-, CP-, and CPT-violating}
probabilities in neutrino-oscillation experiments)
would, in particular, be observed with high-energy neutrinos,
where mass effects can be neglected.

For a possible explanation of a superluminal muon-neutrino
velocity, the previous
FPS model~\cite{KlinkhamerVolovikIJMPA2005} will be modified
in two respects:
\begin{enumerate}
  \item
The hypercharge-based \emph{Ansatz} of Fermi-point (FP) splittings
will be replaced by a pure-neutrino \emph{Ansatz}.
  \item
An additional momentum dependence of the FP splittings
will be introduced in order to
allow for neutrino group and phase velocities different from $c$.
\end{enumerate}
But, first, we must explain what FPS is about in the present context.

Without claims to generality, we consider a particular model
for the $3\times 15$ Weyl fermions of the standard model (SM)
with three additional right-handed neutrinos
[singlet under $SU(2)$ and $SU(3)$ and with zero hypercharge $Y$],
where all masses are neglected (see below).
The dispersion relations of
these $3\times 16$ Weyl fermions [chirality $\chi_{a}^{(f)}=\pm 1$]
are given by
\beq\label{eq:disp-rel}
\Big( E_{a}^{(f)}(p)\Big)^2=\Big( c\,p+ b_{0,\,a}^{(f)}\Big)^2\,,
\eeq
with species label $a \in \{1, \ldots, 16\}$,
family label $f \in \{1, 2, 3\}$,
three-momentum norm $p\equiv |\vec{p}|$, and timelike
FPS parameters $b_{0,\,a}^{(f)}$ to be specified shortly.
Further details can be found in Secs.~5 and 6 of
Ref.~\cite{KlinkhamerVolovikIJMPA2005}.
In the notation of that article,
the species label $a=7$ corresponds to the left-handed
neutrino $\nu_{L}$ and $a=16$ to the right-handed one
$\nu_{R}$, but, here, we will just write $a=\nu_{L}$ and $a=\nu_{R}$,
respectively.

Now, turn to the first modification of the previous FPS model.
Specifically, we propose the following \emph{Ansatz}
for the timelike Fermi-point splittings
entering the dispersion relations \eqref{eq:disp-rel}:%
\bsubeqs\label{eq:b0-Ansatz}
\beqa\label{eq:b0-Ansatz-non-nu}
b_{0,\,a}^{(f)}    &=& 0\,,\quad\;\;\text{for}\;\;
a \ne \nu_{L} \;\wedge\; a \ne \nu_{R}\,,\\[2mm]
\label{eq:b0-Ansatz-nu-L1}
b_{0,\,\nu_{L}}^{(1)}(p) &=& -\xi\,B(p)\,,\\[2mm]
\label{eq:b0-Ansatz-nu-L2}
b_{0,\,\nu_{L}}^{(2)}(p) &=& B(p)\,,\\[2mm]
\label{eq:b0-Ansatz-nu-L3}
b_{0,\,\nu_{L}}^{(3)}(p) &=& -(1-\xi)\,B(p)\,,\\[2mm]
\label{eq:b0-Ansatz-nu-Rf}
b_{0,\,\nu_{R}}^{(f)}(p) &=& -b_{0,\,\nu_{L}}^{(f)}(p)\,,
\eeqa
\esubeqs
with an arbitrary parameter $\xi\in\mathbb{R}$
and a function $B(p)$ to be given later.

In addition to these Fermi-point splittings,\footnote{As
noted in Ref.~\cite{KlinkhamerVolovikIJMPA2005}, there are
really Fermi \emph{surfaces} $E(\vec{p})=0$
if $b_{0,\,a}^{(f)}<0$ in \eqref{eq:disp-rel},
but, for simplicity, we continue to speak about
Fermi \emph{point} splitting. With timelike splittings
as in \eqref{eq:disp-rel}, the propagation of the neutrinos
remains isotropic, whereas spacelike splittings lead to
anisotropy~\cite[(a)]{KlinkhamerJETPL2004-IJMPA2006}.
Remark also that the FPS mechanism, just like the Higgs mechanism,
is associated with spontaneous breaking of gauge symmetries.
As to the dynamical origin of FPS, it could be
an re-entrance effect
(cf. Refs.~\cite{KlinkhamerVolovikIJMPA2005,Klinkhamer2006-review}),
where the intrinsic Lorentz violation at ultrahigh energy
reappears at ultralow energy (Lorentz invariance having emerged
at the intermediate energies).}
we need to specify the corresponding mixing angles and phases
between the interaction (flavor) and propagation states.
We choose them all to be zero.\footnote{Having zero mixing angles
and an FPS parameter $\xi=0$
suppresses the leakage of LV from the neutrino sector
(with $f=2$ and $f=3$) to the electron ($f=1$) sector,
e.g., via quantum-loop corrections to the electron
propagator (cf. Ref.~\cite{MocioiuPospelov2002}).
In addition, zero mixing angles
eliminate new LV contributions to neutrino
oscillations~\cite{KlinkhamerJETPL2004-IJMPA2006,
KlinkhamerPRD2005-PRD2006}.} In the notation of
Ref.~\cite[(b)]{KlinkhamerPRD2005-PRD2006},
the mixing angles and phases are:
\bsubeqs\label{eq:mixing-Ansatz}
\beqa
\chi_{32}&=& \chi_{21}=\chi_{13}=0\,,
\\[2mm]
\omega&=& \alpha=\beta=0\,.
\eeqa
\esubeqs

The crucial property of \emph{Ansatz} \eqref{eq:b0-Ansatz}
is that the induced Chern--Simons-like term
of the photon~\cite{CarrollFieldJackiw1990} vanishes exactly,
as one contribution the purely timelike vector $k_{\mu}^\text{(CS)}$
is determined by the following sum~\cite{KlinkhamerVolovikIJMPA2005}:
\beq\label{eq:Deltak0CS}
\Delta k_{0}^\text{(CS)}\propto
\sum_{a,\,f}\;\chi_{a}^{(f)}\;b_{0,\,a}^{(f)}\;\Big(Y_{a}^{(f)}\Big)^2\,,
\eeq
and another contribution by the same sum
with the hypercharge $Y$ replaced by the isospin $I_3$.
For \emph{Ansatz} \eqref{eq:b0-Ansatz},
only $a=\nu_{L}$ enters the sum and there is a trivial cancellation
between the three different families. In this way,
the proposed FPS model does not ruin the photon propagation properties,
which are tightly constrained by
experiment~\cite{CarrollFieldJackiw1990}.

Two final remarks on the above \emph{Ansatz}.
First, the choice \eqref{eq:b0-Ansatz-nu-Rf} is not unique.
The labels $(f)$ on the right-hand side of \eqref{eq:b0-Ansatz-nu-Rf}
can, for example, be replaced by $(f+1)$, identifying
$(4) \equiv (1)$.
Second, following Ref.~\cite{KlinkhamerVolovikIJMPA2005},
a further contribution $Y_{a}\,\delta b_{0}^{(f)}$,
for constants $|\delta b_{0}^{(f)}|\ll \text{eV}$,
may be added to the values in \eqref{eq:b0-Ansatz}.

Next, turn to the second modification of the previous FPS model.
Specifically, we introduce a nontrivial momentum dependence
in \eqref{eq:b0-Ansatz} by taking
\beq\label{eq:Bfunction}
B(p)= B(0)+\frac{c\,p\;E_\text{vac,\,low}}{c\,p+E_\text{vac,\,high}}\;,
\eeq
with $0<E_\text{vac,\,low}\ll E_\text{vac,\,high}$
and a constant $B(0)$ assumed to be in the sub--eV range.
At this time, \eqref{eq:Bfunction} is simply a working hypothesis,
but it is possible to imagine some type of (nonlocal?)
matter--vacuum interaction to be operative, characterized by `loose' and
`tight' energy scales $E_\text{vac,\,low}$
and $E_\text{vac,\,high}$, respectively.
The energy scales $E_\text{vac,\,low}$ and $E_\text{vac,\,high}$
are, most likely, of the same order as or larger than
the electroweak scale $E_\text{ew}\sim \text{TeV}$.

This completes the definition of the model and we now discuss
some consequences for the propagation of the 48 Weyl fermions.
The dispersion relations \eqref{eq:disp-rel}
for the FPS \textit{Ansatz} \eqref{eq:b0-Ansatz}
imply that all non-neutrino particles have
group velocities equal to the speed of light,
\bsubeqs\label{eq:group-vel-all}
\beqa\label{eq:group-vel-non-nu}
\big( v_\text{group} \big)_{a}^{(f)}
&\equiv& \frac{d E_{a}^{(f)}(p)}{d p}=c\,,
\quad\;\;\text{for}\;\;
a \ne \nu_{L} \;\wedge\; a \ne \nu_{R}\,,
\eeqa
while the neutrinos have
\beqa
\label{eq:group-vel-nu-L1}
\big(v_\text{group}\big)_{\nu_{L}}^{(1)}\big/c - 1&=&
-\xi\;\frac{E_\text{vac,\,low}\,E_\text{vac,\,high}}{(c\,p+E_\text{vac,\,high})^2}
\sim -\xi\;\frac{E_\text{vac,\,low}}{E_\text{vac,\,high}}\,,
\quad\text{for}\;\;
c\,p \ll E_\text{vac,\,high}
\,,\\[2mm]
\label{eq:group-vel-nu-L2}
\big(v_\text{group}\big)_{\nu_{L}}^{(2)}\big/c - 1&=&
\frac{E_\text{vac,\,low}\,E_\text{vac,\,high}}{(c\,p+E_\text{vac,\,high})^2}
\sim \frac{E_\text{vac,\,low}}{E_\text{vac,\,high}}
\,,\\[2mm]
\label{eq:group-vel-nu-L3}
\big(v_\text{group}\big)_{\nu_{L}}^{(3)}\big/c - 1&=&
-(1-\xi)\;\frac{E_\text{vac,\,low}\,E_\text{vac,\,high}}{(c\,p+E_\text{vac,\,high})^2}
\sim -(1-\xi)\;\frac{E_\text{vac,\,low}}{E_\text{vac,\,high}}
\,,\\[2mm]
\label{eq:group-vel-nu-Rf}
\big(v_\text{group}\big)_{\nu_{R}}^{(f)} &=&
\big(v_\text{group}\big)_{\nu_{L}}^{(f)}\,,
\eeqa
\esubeqs
where the approximations in \eqref{eq:group-vel-nu-L2} and
\eqref{eq:group-vel-nu-L3}
also hold for $c\,p \ll E_\text{vac,\,high}$.
Let us briefly comment on the expected mass effects
in the above group velocities.
For a typical neutrino mass $m c^2 \lesssim 1\;\text{eV}$ and
neutrino momentum $p\gg m c$, the relative corrections to the
group velocities are of order
$-(1/2)\,m^2\,c^2/p^2 \lesssim -10^{-19}$ for $c\,p\sim 10\;\text{GeV}$
and are negligible for the present discussion
if the order of magnitude is set by \eqref{eq:OPERA-result}.

 From \eqref{eq:disp-rel} and \eqref{eq:b0-Ansatz},
the following phase velocities are obtained:
\bsubeqs\label{eq:phase-vel-all}
\beqa\label{eq:phase-vel-non-nu}
\big( v_\text{phase} \big)_{a}^{(f)}
&\equiv& \frac{E_{a}^{(f)}(p)}{p}=c\,,
\quad\;\;\text{for}\;\;
a \ne \nu_{L} \;\wedge\; a \ne \nu_{R}
\,,\\[2mm]
\label{eq:phase-vel-non-nuL-nuR}
\big(v_\text{phase}\big)_{a}^{(f)} &\sim&
\big(v_\text{group}\big)_{a}^{(f)}\,,
\quad\;\;\;\;\,\text{for}\;\;
a = \nu_{L} \;\vee\; a = \nu_{R}\,,
\eeqa
\esubeqs
where the last approximate result holds for momenta in the range
$|B(0)| \ll c\,p \ll E_\text{vac,\,high}$.

Most importantly, all  48 Weyl fermions have front velocities
from Eqs.~\eqref{eq:disp-rel}, \eqref{eq:b0-Ansatz}, and
\eqref{eq:Bfunction}, which are equal to the speed of light in vacuum:
\beq\label{eq:front-vel-all}
\big(v_\text{front}\big)_{a}^{(f)}\equiv
\lim_{p\to\infty}\,
\big( v_\text{phase} \big)_{a}^{(f)}
=
\lim_{p\to\infty}\,
\left(\frac{E_{a}^{(f)}(p)}{p}\right)=c\,.
\eeq
Following the work of Sommerfeld and Brillouin
(summarized in Ref.~\cite{Brillouin1960}),
it is known that precisely the front velocity is relevant for the
issue of causality. Very qualitatively, one can say that ultimately
information is carried by a `yes/no' or `$1/0$' signal,
which is infinitely sharp (i.e., discrete), hence the relevance of the
$p\to\infty$ limit in \eqref{eq:front-vel-all}.
See also the related discussion in, e.g., Ref.~\cite{Jackson1975}.

This completes the theoretical discussion of the model
and we now turn to the experimental input.
Identifying the label $f=2$ with the family containing the
charged $\mu^{\pm}$ leptons, the OPERA result \eqref{eq:OPERA-result}
with a generalized magnitude \eqref{eq:def-delta-nu-mu}
implies for our new \FPStype~model
with velocities \eqref{eq:group-vel-nu-L2} and
\eqref{eq:phase-vel-non-nuL-nuR}:
\beq\label{eq:EhighElow-ratio}
E_\text{vac,\,low}/E_\text{vac,\,high} \sim
\delta_{\nu_\mu,\,L}^\text{\,(exp)}\,,
\eeq
as long as the relevant muon-type neutrino energy
$c\, p \sim 17\;\text{GeV}$ lies
in the range $|B(0)| \ll c\,p \ll E_\text{vac,\,high}$.
As discussed in the second and third paragraphs
of this Letter, the magnitude on the right-hand side of
\eqref{eq:EhighElow-ratio} may very well be of order
$10^{-19}$ or less.

Another experimental input comes from the
observed neutrino burst from supernova SN1987a~\cite{SN1987a},
interpreted as a burst of electron-type antineutrinos.
Identifying the label $f=1$
of our \FPStype~model with the family containing the
electron and positron, the implication
from \eqref{eq:group-vel-nu-L1} and \eqref{eq:phase-vel-non-nuL-nuR}
depends on the magnitude of \eqref{eq:def-delta-nu-mu}.
Recalling that the model velocities
\eqref{eq:group-vel-nu-L1} and \eqref{eq:phase-vel-non-nuL-nuR}
are independent of the
neutrino energies (provided these energies are less than
$E_\text{vac,\,high}$), the implication is that
\beqa\label{eq:beta-zero}
|\xi| &\lesssim& 10^{-9}/\delta_{\nu_\mu,\,L}^\text{\,(exp)}\,,
\eeqa
which would give $\xi$ close to zero if the reported magnitude
\eqref{eq:OPERA-result} were correct
but $|\xi|$ could be of order 1 (or larger) if
$\delta_{\nu_\mu,\,L}^\text{\,(exp)}$ were of order $10^{-9}$ (or smaller).

There are two predictions from our \FPStype~model,
the first firm (unconditional), the second loose (conditional):
\begin{enumerate}
  \item
If the muon-neutrino velocity is indeed superluminal, then,
according to \eqref{eq:group-vel-all}, \eqref{eq:phase-vel-all},
and \eqref{eq:EhighElow-ratio},
at least one of the other two flavors of neutrinos must
have a subluminal velocity.
  \item
With nonvanishing effective neutrino FPS from \eqref{eq:b0-Ansatz},
\eqref{eq:Bfunction}, and \eqref{eq:EhighElow-ratio},
and assuming small but nonvanishing mixing angles and phases
instead of \eqref{eq:mixing-Ansatz},
there may be significant T-, CP-, and CPT-violating effects
for neutrino oscillations with baselines of order $10^3\;\text{km}$
and energies of order $10\;\text{GeV}$. See
Ref.~\cite[(b)]{KlinkhamerPRD2005-PRD2006} and, in particular,
Fig.~2 of Ref.~\cite{Klinkhamer2006-review},
which compares models with and without FPS.
\end{enumerate}
A neutrino factory~\cite{Geer1997} would be the ideal tool
to investigate these effects.

\section*{ACKNOWLEDGMENTS}
\noindent It is a pleasure to thank
K. Eitel, G.E. Volovik, and D. Zeppenfeld
for helpful discussions.


\end{document}